\newcommand{\be}{\begin{equation}}
\newcommand{\ee}{\end{equation}}

\documentclass{article}
\pdfoutput=1
\usepackage{amsmath}
\usepackage{amsfonts}
\usepackage{braket}
\usepackage{graphicx}
\usepackage[font={scriptsize,it}]{caption}
\usepackage{amssymb}
\usepackage{mathtools}
\usepackage{xcolor}

\begin{document}
\thispagestyle{plain}
\begin{center}
 \LARGE
Geometry of Entanglement
\vspace{0.7cm}
\large 

\textsc{Andrea Prudenziati \footnote{ andrea.prudenziati@gmail.com}}
\vspace{0.4cm}

\textsl{International Institute of Physics,\\ Universidade Federal do Rio Grande do Norte, \\Campus Universitario, Lagoa Nova, Natal-RN 59078-970, Brazil}
\vspace{0.9cm}
\end{center}


\begin{abstract}
In the context of the surface-state correspondence we propose the geodesic curvature of a convex curve as a local measure of factorization of the dual CFT state. Its integral will be interpreted as computing the total bipartite entanglement among degrees of freedom with support on the chosen domain. We will derive results through application of the Gauss-Bonnet theorem and show quantitative agreement with computations using the MERA tensor network and the formalism of entanglement density. 
\end{abstract}

\section{Introduction and motivations}
Since the proposal from Ryu and Takayanagi to compute entanglement entropy holographically \cite{Hubeny:2007xt},\cite{Ryu:2006bv} and \cite{Ryu:2006ef}, an impressive effort has been done to further develop the relation between entanglement and gravity. A few relevant, cherry picked examples, are \cite{Almheiri:2014lwa}, \cite{Faulkner:2013ica}, \cite{Lashkari:2013koa}, \cite{VanRaamsdonk:2010pw} and \cite{Takayanagi:2017knl}.
The results achieved however mostly suffer, and sometimes exploit, the same common issue that the holographic map is non local. Generally speaking it is only sufficiently close to the boundary that we can restore with good  approximation a local map between the entanglement content of a state and the dual geometry. 

A step forward in this direction has been done with the conjectural surface-state correspondence \cite{Miyaji:2015yva}, where CFT states $\ket{\psi}_{\Sigma}$ are associated to generic convex surfaces $\Sigma$ embedded in the holographic space. These states can in principle be produced by unitary non local operations to the boundary state, so that evolution along the radial bulk direction translates into a purely CFT flux. Unfortunately, with perhaps the only notable exception of constant-radius surfaces that we will discuss later on, we do not have a recipe for constructing  $\ket{\psi}_{\Sigma}$. Nonetheless the simple fact that the Ryu-Takayanagi formula applies also in this context still allows to extract information about the entanglement content of these states; moreover, at least in principle, using such surfaces we can restore the quasi local to local identification between entanglement and geometry at any chosen position in the bulk, the non local part having being pushed entirely into the CFT. A geometrical representation of this flux has been proposed in \cite{Nomura:2018kji}.

A different, and apparently unrelated issue in quantum information, is to define measures of entanglement for multipartite states and/or with infinite dimensionality. Entanglement entropy for instance just captures the entanglement content across the boundary of a certain entangling region. For a review on general entanglement measures see for example \cite{Plenio:2007zz}. 

In the present paper we will attempt both a connection and a step forward in the resolution of these two problems. The aim will be to propose an holographic measure of bipartite entanglement, for a generic 2d CFT state $\rho_{\gamma}$ dual to a convex curve $\gamma$, as an integral of a local geometrical quantity on $\gamma$. The process is summarized as follows: given a small interval $\gamma_A\subset\gamma$  with $\gamma_A=\gamma_{A_1}\cup \gamma_{A_2}$, we begin by choosing a quantum information measure $K(\rho_{\gamma_A}\parallel\rho_{\gamma_{A_1}}\otimes\rho_{\gamma_{A_2}})$ of how much the reduced density matrix $\rho_{\gamma_A}=Tr_{\gamma_A^c}\rho_{\gamma}$  is far from being factorized as $\rho_{\gamma_{A_1}}\otimes\rho_{\gamma_{A_2}}$; a discussion on possible choices of $K$ will follow.  $K(\rho_{\gamma_A}\parallel\rho_{\gamma_{A_1}}\otimes\rho_{\gamma_{A_2}})$ then depends, among other things, on the interval size $\Delta\tau$. We define its second derivative in $\Delta\tau$  local factorizability and write it as $J_{K}(\gamma)$, see (\ref{aa}). We will show that, once integrated along $\gamma$, it provides a measure of the total amount of bipartite entanglement of $\rho_{\gamma}$ inside the integral domain of the dual curve. We claim that $J_K(\gamma)$ can be represented holographically, at least for some choices of measure $K$, as local geometrical quantities on $\gamma$, and we will present an example of this. Then, inverting the process, we will pick a very specific quantity, the geodesic curvature of $\gamma$ and, by studying its dual quantum information properties, we will show that it can be interpreted as representing the holographic dual of $J_K(\gamma)$ for some unknown choice of $K$. Following the quantum information argument the integral of the geodesic curvature along some interval on $\gamma$ should then represent an holographic measure of bipartite entanglement of $\rho_{\gamma}$ there.

With the help of the Gauss-Bonnet theorem we will transform this line integral into a surface integral, compute examples and test our physical interpretation against equivalent computations in the MERA tensor network and entanglement density formalisms. 

When the domain of interest coincides with the total curve, a measure of total bipartite entanglement for the state is then available, and explicit cases will also be discussed.

A final note is that throughout the paper we will consider an holographic description of entanglement at the classical level in the bulk coupling constant $G_N$. This will have an impact on the physical interpretation and results of our analysis. Some  discussion on quantum corrections can be found in the last section.

\section{Local factorizability}
We restrict to three bulk and two boundary dimensions. We will work with a generic space-like slice $M_{sl}$ of the bulk space $M^b$ but for a few explicit examples where we pick $M_{sl}=M^b_{t=\text{const}}$.

We first set our problem by defining what do we mean by local factorizability of the state holographically dual to the convex curve. We will consider for the moment only pure states, $\rho_{\gamma}=(\ket{\psi} \bra{\psi})_{\gamma}$, represented by closed and topologically trivial curves $\gamma \subset M_{sl}$. Density matrices will be briefly discussed in the context of a BTZ black hole.

Consider a connected closed and finite region $\gamma_A$ on $\gamma$ of length $2\Delta\tau$ \footnote{The variable $\tau$ will be the proper length along the curve normalized such that $\parallel\partial_{\tau}\gamma\parallel=1$.}. According to the surface-state correspondence we can associate to $\gamma_A$ the reduced density matrix $\rho_{\gamma_A}=Tr_{\gamma_A^c} (\ket{\psi}\bra{\psi})_{\gamma}$. We then divide $\gamma_A$ into two closed and connected intervals of equal length $\Delta\tau$ such that $\gamma_A=\gamma_{A_1}\cup\gamma_{A_2}$, with  intersection point $p=\gamma_{A_1}\cap\gamma_{A_2}$. Now consider any measure $K(\rho\parallel\sigma)$ of distinguishability between the states $\rho$ and $\sigma$ and define the local factorizability of the state $\ket{\psi}_{\gamma}$ in $p$ with respect to the measure $K$, or $J_K(\gamma)|_p^A$, as 
\be \label{aa}
J_K(\gamma)|_p^A\coloneqq \frac{\partial^2}{\partial(\Delta\tau)^2} K(\rho_{\gamma_A}\parallel\rho_{\gamma_{A_1}}\otimes \rho_{\gamma_{A_2}}).
\ee

The above definition is purely quantum mechanical, with the only holographic connection beeing the dependence of $K(\rho_{\gamma_A}\parallel\rho_{\gamma_{A_1}}\otimes \rho_{\gamma_{A_2}})$ on the size $\Delta\tau$ of the intervals on the dual curve. This is analogous, for instance, to the dependence of the entanglement entropy on the size of the entanglement region, which is also a distance at the boundary of the holographic space. Equation (\ref{aa}) can be applied to  generic finite regions on the curve $\gamma$ but in the following we will mostly be interested in the infinitesimal case $\Delta\tau\ll 1$; when it is so we will drop the subscript $A$ and define local factorizability as the leading order term in $\Delta\tau$ of (\ref{aa}).

There are various choices for measures of distance between quantum states, and to some of them holographic dual quantities have been associated in the literature. Examples are the Fisher information metric, Bures distance and relative entropy, see \cite{Alishahiha:2015rta}, \cite{Alishahiha:2017cuk}, \cite{Banerjee:2017qti}, \cite{Blanco:2013joa}, \cite{Lashkari:2015hha} and \cite{MIyaji:2015mia}. If we had knowledge on the way to compute the state $\rho_{\gamma}$ given a generic convex curve $\gamma$ then these proposal would allow an holographic characterization of (\ref{aa}). Unfortunately in general we do not know how to construct the holographic map $\gamma\rightarrow \rho_{\gamma}$ so some alternative path should be followed. The key insight here is the assumed validity of the Ryu-Takayanagi formula in this context. The reason behind this assumption is that the machinery used to prove the Ryu-Takayanagi formula for boundary states goes through, at least in principle, unaffected when picking generic convex surfaces in the holographic space, \cite{Lewkowycz:2013nqa} and \cite{Miyaji:2015yva}. So, even without knowledge on $\rho_{\gamma}$, we can still extract information on its entanglement content by computing the entanglement entropy for any interval $\gamma_A$ of choice, by measuring the length of the geodesic whose endpoints coincide with $\partial\gamma_A$. Based on this let us  choose $K(\rho\parallel\sigma)=S(\rho\parallel\sigma)$, the latter being the relative entropy:
\[
S(\rho\parallel\sigma)\coloneqq Tr(\rho\log\rho-\rho\log\sigma)
\] 
then it is easily verified that
\[
S(\rho_A\parallel \rho_{A_1}\otimes \rho_{A_2})=I(A_1:A_2),
\] 
with $I(A_1:A_2)=S(A_1)+S(A_2)-S(A_1A_2)$ being the mutual information and $S(A)$ the entanglement entropy. As discussed we can evaluate this quantity holographically for the case of interest by considering the Ryu-Takayanagi surfaces anchored to $\gamma_{A_i}$ \footnote{In the paper \cite{Lashkari:2015hha} Fisher information metric was defined as the second order variation  of relative entropy in some deformation parameter $\lambda$ when considering the CFT ground state ($\lambda=0$) and some slightly higher energy state (at finite $\lambda$). If we could generalize the argument here with base state $\rho_A$ and deformed state $\rho_{\gamma_{A_1}}\otimes \rho_{\gamma_{A_2}}$, and express $\lambda$ as $\lambda(\Delta\tau)$, then we would have some holographic representation of $J_F(\gamma)|_p$ with $F$ the Fisher information metric. I thank some unknown referee for related comments}. In the appendix we computed the mutual information around $p$ for small intervals and then applied the double derivative from the definition (\ref{aa}) to obtain, at first order in the interval size $\Delta\tau$  \footnote{We write $\tau(p)=\tau_p$. Moreover note that $\gamma$ here is called $\tilde{\gamma}$ in the appendix.}:
\be \label{lofa}
J_S(\gamma)|_{p}=\partial^2_{\Delta\tau}I|_{p}=\frac{3\Delta\tau}{4G_N}\left(\partial_{\tau}\gamma^{\theta}\partial_{\tau}\gamma^{\sigma}\partial_{\nu} g_{\theta\sigma}k^{\nu}\right)|_{\tau_p}+O(\Delta\tau^3).
\ee
The expression (\ref{lofa}) is interesting for different reasons. First of all it depends on the vector $k^{\mu}$, whose norm is called the geodesic curvature and which is defined as \footnote{ \label{ftn}In the bulk $M^b$ we have $k_c^{\mu}(\gamma)\coloneqq\bigtriangledown_{\tau}\partial_{\tau}\gamma^{\mu}$ whose norm is related to the norms of the projection over the tangent plane to the space-like slice $M_{sl}$ in the chosen point, $k^{\mu}(\gamma)\coloneqq(\bigtriangledown_{\tau}\partial_{\tau}\gamma^{\mu})_{TM_{sl}}$ and the normal part, $k_n^{\mu}(\gamma)\coloneqq(\bigtriangledown_{\tau}\partial_{\tau}\gamma^{\mu})_{NM_{sl}}$ by $k_c=\sqrt{k^2+k_n^2}$. }:
\be\label{gcurv}
 k^{\mu}(\gamma)\coloneqq(\bigtriangledown_{\tau}\partial_{\tau}\gamma^{\mu})_{TM_{sl}}, \;\;\;\;\; \parallel\partial_{\tau}\gamma\parallel=1,
\ee
with the obvious consequence that $k^{\mu}(\gamma=\text{geodesic})=0$ so that $J_S(\gamma=\text{geodesic})=0$. Geometrically we can simply visualize this: if a curve is a geodesic on a finite interval $\gamma_B$, then there  the curve coincides with the Ryu-Takayanagi surface for any interval $\gamma_A\subset\gamma_{B}$, so the holographic mutual information of the dual state trivially vanishes, the state is exactly factorized as $\rho_{\gamma_{A}}= \rho_{\gamma_{A_1}}\otimes \rho_{\gamma_{A_2}}$ and $J_S(\gamma)|_{p\in\gamma_A}=0$. In fact the property that $\rho_{\gamma_{A}}= \rho_{\gamma_{A_1}}\otimes \rho_{\gamma_{A_2}}$ when the dual curve $\gamma$ is a geodesic along some segment means that  $K(\rho_{\gamma_{A}}\parallel\rho_{\gamma_{A_1}}\otimes \rho_{\gamma_{A_2}})=0$ for \emph{any} choice of $K$, which means that local factorizability always vanishes there. This implies that the holographic dual to (\ref{aa}), for whatever choice of measure $K$, should always be proportional to a geometrical quantity bound to vanish when evaluated along a geodesic, for instance some power of the geodesic curvature. 

The double derivative in (\ref{aa}) can roughly be interpreted as the quantum information equivalent of the "geodesic" equation for a generic curve (\ref{gcurv}).
In fact, without the second order derivative in (\ref{aa}), the local factorizability would vanish  as $\sim\Delta \tau^3$ for $\Delta\tau\rightarrow 0$, much more severely then in (\ref{lofa}); one power of $\Delta \tau$ comes from the integration domain and two powers from the geometric property that a generic curve is infinitesimally different from its tangent geodesic in $p$ only at second order in the distance from $p$. Finally, if we consider as an example the Poincar\'e metric of $\text{AdS}_3$ with radius $L$, then 
\be\label{weight}
\partial_{\nu} g_{\theta\sigma}k^{\nu}\sim - 2 \frac{L^2}{z^3} k^z,
\ee
which sets the relative weight of $J_S(\gamma)|_p$ compared to curves with different bulk embedding, depending on the  radial position of $\gamma(p)$.

\section{Definition of $J(\gamma)$}
The scaling result (\ref{weight}) shows that local factorizability with measure the relative entropy (\ref{lofa}) does weight the entanglement between $\rho_{A_1}$ and $\rho_{A_2}$ with a scaling factor of $\sim z^{-3}$. Following the MERA representation of AdS we interpret the AdS radius as some scale of entanglement after a coarse graining procedure has been implemented, see section \ref{meras} with related references. However the MERA tensor network weights entanglement at different length scales equally, so to have agreement with this counting we would like to get rid of the additional factor (\ref{weight}). Analogously we can count entanglement between degrees of freedom at different length scales using the formalism of entanglement density, see section \ref{ed}; also here the qubits pairs are identically weighted independently of their distance. Thus inspired by the holographic computation of local factorizability using relative entropy (\ref{lofa}), we would like to study a "simpler" but similar geometrical quantity: 
\be \label{lofax}
J(\gamma)|_p\coloneqq \frac{L}{4 G_N} k(\gamma)|_{\tau_p}\Delta\tau.
\ee
This definition is purely geometrical in nature but eventually we would like to interpret it as the holographic dual of local factorizability for the state $\ket{\psi}_{\gamma}$ based on some unknown measure  $X$ of state distinguishability, so that $J(\gamma)=J_X(\gamma)$. The definition is with respect to some infinitesimal interval of size $2\Delta\tau$ centred at $\tau_p$.
Above we have set $k(\gamma)$ to be the norm of $k^{\mu}(\gamma)$ from (\ref{gcurv})  which is called the geodesic curvature of $\gamma$. In the following sections we will provide evidence for this claim, for the moment we simply emphasize that this definition retains the following good properties:
\begin{itemize}
\item $J(\gamma)|_p=0$ when $\gamma$ coincides with a geodesic on $\gamma_B \supset \gamma_A$, with $p\in \gamma_A$, which implies (from the discussion on holographic mutual information) that $\rho_{\gamma_{A}}= \rho_{\gamma_{A_1}}\otimes \rho_{\gamma_{A_2}}$ there.
\item It is positive definite in any point $p$ for generic convex curves.
\item It is integrable along $\gamma$ in $\tau_p$ by replacing $\Delta\tau\rightarrow d\tau$. 
\item It is dimensionless.
\end{itemize}
Notice that (\ref{lofax}) differs substantially from (\ref{lofa}) as it is a more "democratic" measure of local factorizability at different $\text{AdS}_3$ length scales, not being weighted by the radial depth in the bulk space. 

Additivity of some measure $E$ is defined as $E(\rho_1\otimes\rho_2)=E(\rho_1)+E(\rho_2)$. In holography we can consider two bulk spaces $M_1$ and $M_2$ dual to the states $\rho_1$ and $\rho_2$ and obtain the bulk space $M$ dual to $\rho=\rho_1\otimes\rho_2$ as the disconnected union $M=M_1\cup M_2$. Then it is trivially verified that
\begin{itemize}
\item $J(\gamma)$ is additive.
\end{itemize}
Finally let us discuss Local Operations (LO), as a good entanglement measure should not increase under LO \footnote{and Classical Communication, although how to describe CC in the holographic setting is quite unclear to me.}. We first choose a CFT unitary transformation $U(\gamma_{A})$ that has domain of application on $\gamma_{A}$ and acts as  $U(\gamma_A)\ket{\psi}_{\gamma}=\ket{\psi}_{\gamma'}$, with $\gamma'-\gamma\coloneqq \delta\gamma $ and $\delta\gamma\neq 0$ only inside $\gamma_A$.  As the boundary conditions of the deformation are $\delta\gamma|_{\partial\gamma_A}=0$, the Ryu-Takayanagi formula tells us that $S(\gamma_A)=S(\gamma'_A)$ so that we do not modify the entanglement entropy of the reduced density matrix $\rho_{\gamma_A}$ under the action of $U(\gamma_{A})$, and we may call the latter LO for the subsystem $\gamma_A$. Our formalism is consistent with this identification because if we apply these LO on $\gamma_{A_1}$ and $\gamma_{A_2}$ they transform $\gamma$ there but keeping fixed the boundary points at $\partial\gamma_{A_1}$ and $\partial\gamma_{A_2}$. Then (for notational simplicity we skip below the subscript $TM_{sl}$)
\begin{multline*}
|k(\gamma')-k(\gamma)|_{\tau_p}=|\parallel\bigtriangledown_{\tau}\partial_{\tau}\gamma'\parallel-\parallel\bigtriangledown_{\tau}
\partial_{\tau}\gamma\parallel|_{\tau_p}\leq \parallel \bigtriangledown_{\tau}\partial_{\tau}\delta\gamma\parallel_{\tau_p}=\\
=\parallel \partial_{\tau}^2 \delta\gamma+\Gamma_{\mu\nu}\partial_{\tau}\delta\gamma_{\mu}\partial_{\tau}\delta\gamma_{\nu}\parallel_{\tau_p}=0,
\end{multline*}
 where we used the fact that the derivatives are expressed as incremental differences in $\Delta\tau\rightarrow d\tau$ of $\delta\gamma$, which means evaluated at the boundary points of $\partial\gamma_{A_1}$ and $\partial\gamma_{A_2}$. We conclude that
\begin{itemize}
\item $J(\gamma)$ does not vary under LO \footnote{For finite intervals we should either require additional boundary conditions on $\delta\gamma$ at $\partial\gamma_A$ as part of the definition of LO, if we want to preserve $J(\gamma)$, or alternatively accept that local factorizability can increase under the more general definition of LO. }.
\end{itemize}

\section{Applications}
\begin{figure}
    \centering
    \includegraphics[width=\textwidth]{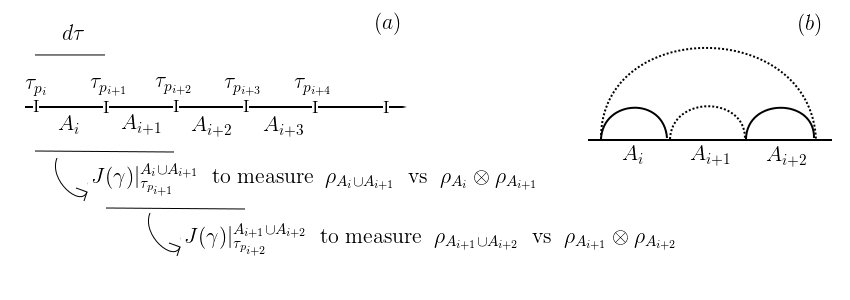}
    \caption{Part (a) is a graphic representation of the (discretized) integral on $\tau_p$ of $J(\gamma)$, while part (b) shows the vanishing of (bulk classical) mutual information for non contiguous intervals.}
    \label{figurafact}
\end{figure}

To study the properties of $J(\gamma)$ we will consider some generic state $\ket{\psi}_{\gamma}$ and integrate (\ref{lofax}) in $\tau_p$ along some region $\gamma_A$, having done the substitution $\Delta\tau \rightarrow d\tau$:
\be \label{intlf}
J(\gamma_A)\coloneqq\int_{\gamma_A} J(\gamma)|_{\tau_p}=\frac{L}{4 G_N} \int_{\gamma_A}k \; d\tau.
\ee
 In this way $\gamma_A$ is divided into infinitely many subregions $A_i$ of size $d\tau$ with boundary points $\tau_{p_{i}}$ and $\tau_{p_{i+1}}$. The integral on $\tau_p$ will then add up all the local factorizability values measuring how much the state $\ket{\psi}_{\gamma}$ is far from being factorized across any couple of infinitesimal contiguous subregions $A_i,A_{i+1}$, see part $(a)$ of figure \ref{figurafact}. 

Two observations generalize this result: first of all if we want to describe factorization of the state between two non-contiguous subregions at the \emph{classical} level in the bulk coupling constant $G_N$, for instance $A_i,A_{i+2}$ by measuring the mutual information $I(A_i:A_{i+2})$, this is trivially zero. The reason being that the configuration of the two disconnected geodesics with boundary $\partial A_i$ and $\partial A_{i+2}$ is favoured in this case (of regions of identical size) over the case of the two disconnected geodesics with boundary points $\partial (A_i \cup A_{i+1} \cup A_{i+2})$ and $\partial A_{i+1}$, see part $(b)$ of figure \ref{figurafact}. So factorization along non contiguous regions is described holographically by quantum corrections in the bulk and we expect this to hold true even when using different entanglement measures than the relative entropy. Because of this we should not worry about the distinction between a state that is factorized only between all the contiguous couples of $A_i$s and one that is factorized along \emph{any} couple, if we stick with a classical bulk regime. This will be important when we will compare results with the entanglement density formalism in section \ref{ed}

Moreover rescaling $\tau\rightarrow \lambda\tau$ changes the interval size of the subregions $A_i$ as $d\tau\rightarrow \lambda d\tau$, so that the above discussion holds true for two parties factorization of any two unions of $\lambda$ contiguous subregions, $\cup_{i=j}^{j+\lambda} A_i$ and $\cup_{i=j+\lambda+1}^{j+2\lambda+1} A_i$. This observation will be relevant for the comparison with results from the MERA tensor network in section \ref{meras}.

Given the above discussion the integral (\ref{intlf}) is then interpreted as a measure of the total amount of two parties factorization of $\ket{\psi}_{\gamma}$, alternatively referred to as bipartite entanglement, evaluated over all couples of regions with support on $\gamma_A$. Seen from a lattice point of view this is a sum of entanglement from all the couples of degrees of freedom. The holographic dual description is just the measure of how much the curve differs from being a geodesic along $\gamma_A$.  

What we cannot deduce however is for $J(\gamma)$ to be a measure of \emph{total} factorization, that is how much $\rho_{\cup_i A_i}$ differs from $\rho_{A_1}\otimes\rho_{A_2}\otimes \dots \otimes \rho_{A_n}$. Even in a three parties state two parties entanglement is not enough to reconstruct the general three parties entanglement; a simple example of this is the three qubits entangled GHZ state $\ket{GHZ}=\frac{1}{\sqrt{2}}(\ket{000}+\ket{111})$ whose reduced density matrix by tracing out any of the three qubits is \emph{not} entangled in the remaining two qubits. For general multipartite states the situation becomes even more involved so that it is indeed hopeless to imply that a state is totally factorized, based on two parties factorization alone, and in particular does not make (\ref{intlf}) a measure of the total amount of entanglement of the state.

\subsection{Entanglement below a geodesic and the Gauss-Bonnet theorem}
We compute here the difference in $J$ between two states that share the same smooth curve $\gamma^c$ but for a certain region $A$ (for simplicity here connected but this can be straightforwardly generalized) where one state follows a smooth curve $\gamma_A$ and the other $\tilde{\gamma}_A$. For generality we also include the jump angles as in figure \ref{figGA}, that are nothing but local divergences of the geodesic curvature:
\begin{figure}
    \centering
    \includegraphics[width=\textwidth]{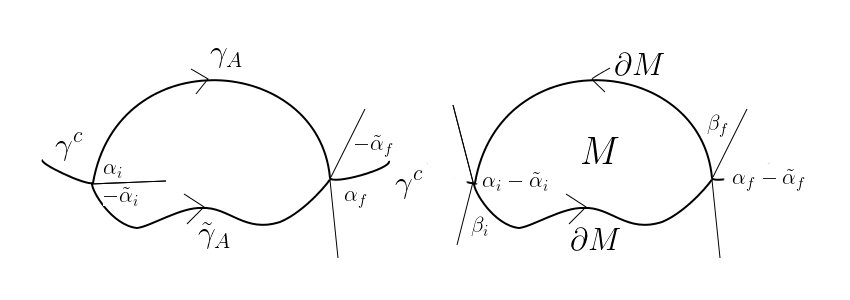}
    \caption{Difference in integrated local factorizability between two curves, jump angles are taken in the counter clockwise direction.}
    \label{figGA}
\end{figure}
\begin{multline}\label{temp}
\frac{4G_N}{L}\left( J(\gamma^c\cup\tilde{\gamma}_A)-J(\gamma^c\cup\gamma_A)\right)=\tilde{\alpha}_i+\tilde{\alpha}_f+\int_{\gamma^c\cup\tilde{\gamma}_A}
\hspace{-0.5cm}k-
\left(\alpha_i+\alpha_f+\int_{\gamma^c\cup\gamma_A}\hspace{-0.5cm}k \;\;\;\right)=\\ =\beta_i+\beta_f-2\pi+\int_{\delta M}\hspace{-0.1cm}k,
\end{multline}
where we have denoted $\partial M=(-\gamma_A)\cup\tilde{\gamma}_A $, while $\beta=\pi-(\alpha-\tilde{\alpha})$ are the jump angles of the piecewise smooth curve $\partial M$. Following figure \ref{figGA} then  
\be
J(\gamma^c\cup\tilde{\gamma}_A)-J(\gamma^c\cup\gamma_A)=J(\partial M)-2\pi \frac{L}{4G_N}.
\ee
An important result that we will use extensively in the following is the Gauss-Bonnet theorem that states:
\be 
\int_{\partial M}\hspace{-0.1cm}k+\sum_i \alpha_i +\int_M \frac{R}{2}=2\pi \chi,
\ee
where $\chi$ is the topological Euler number of $M$ and $R$ is the Ricci scalar. It then follows that \footnote{The formula below applies also to the case of $A$ being made by $n$ disconnected intervals, with $M$ being the corresponding $n$ disconnected regions, simply replacing $2\pi(\chi-1)\rightarrow 2\pi(\chi-n)=4\pi(1-n)$, because of the additional $n$ factors of $2\pi$ from (\ref{temp}) and the Euler number definition.}:
\be \label{diff}
J(\gamma^c\cup\tilde{\gamma}_A)-J(\gamma^c\cup\gamma_A)=\frac{L}{4G_N}\left(-\int_M \frac{R}{2}+2\pi(\chi-1)\right)=-\frac{L}{4G_N}\int_M \frac{R}{2}.
\ee
This interesting result is interpreted to measure the difference in  total bipartite entanglement between the states $\ket{\psi}_{\gamma^c\cup\tilde{\gamma}_A}$ and $\ket{\psi}_{\gamma^c\cup\gamma_A}$. 

As a nice application of this formula let us consider the case where $\gamma_A$ is a geodesic with boundary conditions $\partial \gamma_A=\partial \tilde{\gamma}_A$. In this case the state $\ket{\psi}_{\gamma^c\cup\gamma_A}$ is completely factorized on $\gamma_A$ so no local entanglement should be present there and indeed  $\int_{\gamma_A}\hspace{-0.2cm}k=0$ in this case. Then (\ref{diff}) transforms into
\be \label{diff2}
J^{\text{tot}}(\tilde{\gamma}_A)\coloneqq J(\gamma^c\cup\tilde{\gamma}_A)-J(\gamma^c\cup\gamma_A)|_{\gamma_A=\text{RT\;curve}}=-\frac{L}{4G_N}\int_M \frac{R}{2},
\ee
In this case we indeed interpret $J_{\tilde{\gamma}_A}^{\text{tot}}$ as a measure of the total amount of two parties local entanglement on $\tilde{\gamma}_A$. 

We can compute (\ref{diff2}) for the simple case of $\tilde{\gamma}$ being the infinitesimal boundary cutoff state $z=\epsilon$ in the Poincar\'e $\text{AdS}_3$ coordinates at $t=\text{const}$ of subsection \ref{poin}, and $\tilde{\gamma}_A$ a sub region of length $l$. The geodesic $\gamma_A$ then is a semicircle and $M$ the region inside, which makes the computation and the result remarkably simple:
\begin{multline}\label{res}
J^{\text{tot}}(z={\epsilon},\; x\in[-l/2,l/2])=-\frac{L}{4G_N}\int_M \frac{R}{2} =\\=\frac{L}{4G_N}\int_{\delta}^{l/2}\hspace{-0.3cm}dr\int_{\epsilon/r}^{\pi-\epsilon/r}\hspace{-0.6cm}d\phi\frac{L^2}{r\sin^2\phi} \frac{1}{L^2}\xrightarrow[\text{$\epsilon,\delta\rightarrow 0$}]{} \frac{c}{6}\frac{l}{\epsilon},
\end{multline}
where we set the temporary cutoff $\delta$ for the size of $A$ to be $l\gg\delta\gg \epsilon$ and we have used the famous relation $\frac{1}{G_N}=\frac{2c}{3L}$.  

Interesting is also the computation for finite boundary cutoff $z=z_c$; as the geodesics now is no longer a half circle we find easier to switch back to the line integral formalism. We parametrize the $z=z_c$ curve as $x(\tau)=\frac{z_c}{L} \tau +\text{const}$ and compute the geodesic curvature along to be simply $k=1/L$, to obtain:
\begin{multline}\label{res2}
J^{\text{tot}}(z=z_c,\; x\in[-l/2,l/2])=\frac{L}{4G_N}\left(\beta_i+\beta_f-2\pi+\int_{\delta M}\hspace{-0.1cm}k \right)=\\=\frac{L}{4G_N} \int_{\tau_i}^{\tau_f}d\tau\frac{1}{L}=\frac{c}{6}\frac{l}{z_c},
\end{multline}
which is a deceivingly simple generalization of (\ref{res}). 

In the following two sub sections we will test our conjectural interpretation of (\ref{diff2}) by comparing the results (\ref{res}) and (\ref{res2}) with two different approaches for estimating bipartite entanglement inside some boundary $\text{AdS}_3$ region: the MERA tensor network and entanglement density formalism.

\subsection{Comparison with MERA}\label{meras}
MERA \cite{Vidal:2007hda} is a tensor network that represents some given state, for instance an $n$ qubits chain, by an iterative application of tensor disentanglers acting  on nearby qubits followed by coarse graining. This layered network has been proposed to represent a discretized version of the holographic bulk space \cite{Swingle:2009bg}, with the iterative direction being interpreted as the additional bulk dimension. As each disentangling operation at a given scale $u$ acts by killing entanglement between two coarse grained degrees of freedom at the same scale $u$, by counting the minimal total number of such operators in the network surrounding the entangling region $A$,  we should estimate the total entanglement of $A$ with its complementary, which is the entanglement entropy of $A$. And indeed if we start with $A$ made of $\frac{l}{\epsilon}$ qubits (here we imagine a qubit for each lattice site of size $\epsilon$), at the level $u$ this number has been reduced by coarse graining to be $\frac{l}{\epsilon}2^{-u}$, $u\geq 0$. The number of disentanglers is then half of the number of qubits at level $u$, where $u$ is related to the $z$ radius in Poincar\'e coordinates by $z=\epsilon 2^u$. Thus the total number of disentanglers surrounding $A$ can be represented as a "geodesic" curve in the MERA network, attached to the boundary of $A$, that counts the entanglement as the total number of intersections with disentangler tensors \footnote{the factor of $2$ comes from the two sides of the geodesic while the factor $1/2$ from the relative counting of disentanglers compared to the level}: 
\be
\#\text{disent.}\approx 2\int_{0}^{\log_2 l/\epsilon}\hspace{-0.5cm}du\frac{1}{2}=\frac{1}{\log 2}\log\left(\frac{l}{\epsilon}\right).
\ee
This indeed coincides with the Ryu-Takayanagi formula if we weight the entanglement contribution of each tensor contraction by a factor $\log 2\frac{c}{3}$. 

This simple interpretation of entanglement across $A$ can be refined to obtain (\ref{res}). We proposed this formula as a measure of the total amount of two parties entanglement among regions of any (identical) size living in $A$, that from the MERA perspective should then correspond to the \emph{total} number of disentanglers below the Ryu-Takayanagi curve in the network. Including the $\log 2\frac{c}{3}$ weight from above this  is given by :
\be \label{mera}
\text{tot.\;2-ent.\;in}\;A\approx \log 2\frac{c}{3}\int_{0}^{\log_2 l/\epsilon}\hspace{-0.3cm}du \frac{1}{2}\frac{l}{\epsilon} 2^{-u}=\frac{c}{6}\frac{l}{\epsilon}+O(1),
\ee
that again agrees with (\ref{res}).

The generalization for a curve at $z=z_c$ is trivially achieved by starting at level $u=0$ from $\frac{l}{z_c}$ qubits, and so replacing $\epsilon\rightarrow z_c$ in equation (\ref{mera}) to nicely match (\ref{res2}).

\subsection{Comparison with Entanglement Density}\label{ed}
Entanglement density $n(\delta,\xi)$ was first introduced in \cite{Nozaki:2013wia} and is defined in a lattice theory as the number of entangled pairs centred at $\xi$ and separated by a distance $\delta$. A simple calculation performed in \cite{Nozaki:2013wia} shows that the entanglement entropy is related to the entanglement density as $n(\delta,\xi)=\frac{1}{8}\partial^2_{\xi}S(\delta,\xi)-\frac{1}{2}\partial^2_{\delta}S(\delta,\xi)$. In the specific case of the $\text{AdS}_3$ boundary state $\gamma=\{z=\epsilon\}$ entanglement entropy has the simple value $S(\delta,\xi)=S(\delta)=\frac{c}{3}\log(\frac{\delta}{\epsilon})$, so that
\be
n(\delta)=\frac{c}{6}\frac{1}{\delta^2}. 
\ee
Note that the radial cutoff decouples and the entanglement density is cutoff independent.

To compute the total two parties entanglement below the Ryu-Takayanagi curve for a boundary entangling region of size $l$ ( for simplicity centred at the origin), we should then count all such entangled pairs that fit inside the entangling region, with the cutoff $\epsilon$ introducing a minimal distance for each couple. This is easily computed as:
\be\label{provaentdens}
\text{tot.\;2-ent.\;in}\;A\approx 2\int_0^{l/2-\epsilon/2}\hspace{-0.5cm}d\xi\int_{\epsilon}^{l-2\xi} \hspace{-0.4cm}d\delta\;n(\delta)=\frac{c}{6}\frac{l}{\epsilon}-\frac{c}{6}\log\left(\frac{l}{\epsilon}\right)+O(1).
\ee 
This result agrees with (\ref{res}) in the leading term thus supporting our interpretation of the integral (\ref{diff2}). The additional subleading term in (\ref{provaentdens}), that is nothing but minus a half of the corresponding entanglement entropy, is probably due to some under counting of entangled pairs near the boundary of the Ryu-Takayanagi surface, that this simple evaluation missed.

More interesting is the comparison for a curve at finite cutoff $z_c$, because the Ryu-Takayanagi formula in this case predicts the entanglement entropy to be \footnote{It is perhaps interesting to note that the form (\ref{entdensz}) for entanglement entropy can be reduced to the usual logarithmic scaling if we introduce an effective cutoff $\tilde{z}_c(l,z_c)= \frac{z_c l^2}{l^2 - z_c^2}$ such that $S(l)_{\tilde{z}_c(l,z_c)}=\frac{c}{3}\text{log}\left(\frac{l}{z_c}\right)$. This may be relevant in the ensuing discussion at the end of this section.}
\begin{equation}\label{entdensz}
S(l)_{z_c}=\frac{c}{3}\text{ArcSech}\left(\frac{2z_c}{\sqrt{l^2+4z_c^2}}\right).
\end{equation}
The entanglement density can be defined also for finite cutoff, generalizing the original computation in \cite{Nozaki:2013wia}, to obtain $n(\delta,\xi-z_c/2,z_c)+n(\delta,\xi+z_c/2,z_c)=\frac{1}{4}\partial^2_{\xi}S(\delta,\xi)_{z_c}-\partial^2_{\delta}S(\delta,\xi)_{z_c}$. Translational invariance simplifies the relation to $n(\delta,z_c)=-\frac{1}{2}\partial^2_{\delta}S(\delta)_{z_c}$ and using (\ref{entdensz}) we have
\begin{equation}
n(\delta,z_c) = \frac{c \delta}{6 \left(4 z_c^2 + \delta^2\right)^{\frac{3}{2}}},
\end{equation}
with the cutoff now appearing inside the entanglement density. Repeating the computation (\ref{provaentdens}) we then obtain
\begin{multline}\label{provaf}
\text{tot.\;2-ent.\;in}\;A|_{z_c}\approx 2\int_0^{l/2-z_c/2}\hspace{-0.5cm}d\xi\int_{z_c}^{l-2\xi} \hspace{-0.6cm}d\delta \;n(\delta,z_c) =\\= \frac{1}{\sqrt{5}}\frac{c}{6}\frac{l}{z_c}-\sqrt{5}\frac{c}{{6}}\log\left(\frac{l+\sqrt{l^2+4z^2}}{z_c}\right)+O(1).
\end{multline}
This time we do not have agreement with (\ref{res2}) in the leading term because of the multiplying factor of $1/\sqrt{5}$. We notice a couple of points: first even with this slight mismatch it is indeed remarkable that the scaling remains linear in $l/z_c$, considering all the path followed. From the CFT point of view this last computation is probably oversimplified, the state dual to the finite $z=z_c$ curve having being recently identified with the vacuum of the $\text{T}\bar{\text{T}}-$deformed theory with deformation parameter $\mu(z_c)$, \cite{Chen:2019mis} and \cite{McGough:2016lol}. Thus we expect to need a more refined counting of entanglement couples to match our proposal, probably a slightly different definition of entanglement density starting from entanglement entropy, which is the link between the CFT interpretation and the holographic data. It is perhaps important to stress out that, as in the more general case of surface-state correspondence, the Ryu-Takayanagi formula is still believed to hold for T$\bar{\text{T}}$-deformed theories, from both the holographic replica trick argument of \cite{Lewkowycz:2013nqa} and explicit computations, \cite{Chen:2018eqk} and \cite{Donnelly:2018bef}.

\subsection{Total (two parties) entanglement of a state}
Following these ideas we are naturally led to consider equation (\ref{diff2}) for a space filling region. What happens to the geodesic $\gamma_A$, that bounds $M$ together with $\tilde{\gamma}_A$, depends on the bulk geometry. For example it vanishes in $AdS_3$ but wraps the horizon for a BTZ black hole. The simple geometrical computation of $J_{\tilde{\gamma}}$ now upgrades its physical interpretation to a measure of the total two parties entanglement of the state $\ket{\psi}_{\tilde{\gamma}}$. 

\subsubsection*{$\text{AdS}_3$ in global coordinates}
We consider here $\tilde{\gamma}$ to be the curve at constant $\rho=\rho_{UV}\gg 1$ in global $\text{AdS}_3$  coordinates at $t=\text{const}$:
\[
ds^2=L^2\left(d\rho^2+\sinh^2\rho \;d\phi^2 \right).
\]
The corresponding geodesic $\gamma_A$ vanishes as $\tilde{\gamma}_A$ extends to include the full $\phi$-domain, so that $J_{\tilde{\gamma}}$ becomes a measure of total two parties entanglement for the state $\ket{\psi}_{\rho=\rho_{UV}}$. We can compute the integral (\ref{diff2}) to obtain:
\be 
J^{\text{tot}}(\rho=\rho_{UV})=-\frac{L}{4G_N}\int_{0}^{\rho_{UV}}\hspace{-0.3cm}d\rho\int_0^{2\pi}\hspace{-0.2cm}d\phi\sqrt{g}\frac{R}{2}=\frac{\pi L}{2G_N}\int_{0}^{\rho_{UV}}\hspace{-0.3cm}d\rho\sinh\rho  
\ee 
\[
\xrightarrow[\text{$\rho_{UV}\rightarrow \infty$}]{}  \frac{\pi L}{4G_N}e^{\rho_{UV}}=\frac{c}{6}\pi e^{\rho_{UV}}.
\]
Note that this is the same result we would get by simply computing 
\[
\frac{\text{length}(\partial \text{AdS}_3|{t=0})}{4G_N}=\frac{1}{4G_N}\int_{0}^{2\pi}\hspace{-0.2cm}d\phi\sqrt{g_{\phi\phi}}\xrightarrow[\text{$\rho_{UV}\rightarrow \infty$}]{} \frac{\pi L}{4G_N}e^{\rho_{UV}}=\frac{c}{6}\pi e^{\rho_{UV}}\approx N_{dof}.
\]
The last equivalence with $N_{dof}$ comes from the idea that a gravitational theory contains as many physical degrees of freedom as the size of its boundary in Plank units. Indeed also a quantity that scales as $\sim c\frac{l}{\epsilon}$ for a finite region $l$ and cutoff $\epsilon$ should be seen as proportional to the number of local degrees of freedom. So in general we may argue that
\be
J^{\text{tot}}(\tilde{\gamma}_A)\approx \text{total\;2-entanglement\;in\;}\tilde{\gamma}_A\approx N_{dof} \;\text{in}\;\tilde{\gamma}_A
\ee
Perhaps it would be interesting to exploit further this relation in the context of black holes and entanglement production by Hawking radiation, which is indeed an example of two parties entanglement.  

\subsubsection*{$\text{AdS}_3$ in Poincar\'e coordinates}\label{poin}
Similarly to the previous case we still consider $\text{AdS}_3$ at $t=\text{const}$ but this time the cutoff is in Poincar\'e coordinates \footnote{That covers only one half of the total $\text{AdS}_3$ space.}:
\[
ds^2=\frac{L^2}{z^2}\left(dz^2+dx^2\right)
\]
and $\tilde{\gamma}$ corresponds to $z=\epsilon\ll 1$. Then introducing a large IR cutoff $C\gg1$ on the $x$ coordinate:
\be 
J^{\text{tot}}(z=\epsilon)=-\frac{L}{4G_N}\int_{\epsilon}^{\infty} \hspace{-0.2cm}dz\int_{C/2}^{C/2}\hspace{-0.2cm} dx\sqrt{g}\frac{R}{2}=\frac{C L}{4 G_N}\int_{\epsilon}^{\infty}\hspace{-0.2cm} dz\frac{1}{z^2}=\frac{C L}{4 G_N}\frac{1}{\epsilon}
\ee 

\subsubsection*{BTZ black hole}
The final simple example is a BTZ black hole at constant time:
\[  
ds^2=\frac{L^2r^2}{(r^2-r_-^2)(r^2-r_+^2)}dr^2+r^2d\phi^2.
\]
and for simplicity we pick zero angular momentum or $r_-=0$. Here the CFT states dual to generic curves are not pure, so all the physical interpretation we have attached loses much of its meaning, starting with the discussion on mutual information and factorizability, as the procedure of constructing a reduced density matrix mixes up entanglement effects with the entropy of the original density matrix we have started with. Nonetheless it is easy to compute $J^{\text{tot}}$ also in this case so we will just do it.

The geodesic $\gamma_A$ now wraps the horizon at $r=r_+$ for $A$ space filling. So our integral (\ref{diff2}) will span all the region in between the horizon and the curve $\tilde{\gamma}$ that we choose to be at $r=r_{UV}>>1$. We obtain:
\be 
J^{\text{tot}}(r=r_{UV})=-\frac{L}{4G_N}\int_{r_+}^{r_{UV}} \hspace{-0.3cm}dr\int_0^{2\pi}\hspace{-0.2cm} d\phi\sqrt{g}\frac{R}{2}=\frac{\pi}{2G_N}\int_{r_+}^{r_{UV}} \hspace{-0.3cm}dr \frac{r}{\sqrt{r^2-r_+^2}}
\ee
\[
=  \frac{\pi}{2G_N}\sqrt{r_{UV}^2-r_{+}^2}.
\]

\section{Conclusions and future work}
We have discussed a geometric local measure of factorization of CFT states dual to some convex curve as the geodesic curvature of such a curve. This geometric measure $J(\gamma)$ is argued to have its dual counterpart as the second derivative in the interval size of some quantum information distance between a two-parties entangled state and its factorized version. Inspiration has come from the holographic computation of mutual information for small intervals, by using the relative entropy as a distance between states. Quantum properties of a "good" entanglement measure have been studied, and the physical interpretation of the integral $J(\gamma_A)=\int_{\gamma_A} J(\gamma)|_{\tau_p}$ as a measure of bipartite entanglement within $\gamma_A$ discussed. The use of the Gauss-Bonnet theorem permits a simple computation of such a quantity and the case of a boundary state has been compared with results from the entanglement density and MERA tensor network formalisms. As the physical interpretation of the latter is quite clear, agreement with the result confirms our derivation. Finally some simple examples of total bipartite entanglement for the full state have been computed, using the boundary curve of $\text{AdS}_3$ at two different cutoffs and the BTZ black hole, where the entanglement interpretation becomes  less transparent.

There are some main points that deserve further study in the future:

\subsubsection*{Bulk reconstruction}
In this paper we pushed a little bit forward the geometry-to-entanglement map by identifying local factorizability of generic CFT states dual to convex curves with the local geodesic curvature of the curve. This association has the merit of being local, but the demerit of being based on some unknown state-measure and non-local CFT transformation from the usual boundary state to our curve state. In other words we have moved all the non-locality to the CFT side. I believe this to be a good starting point to better understand the bulk reconstruction problem, see for instance \cite{Dong:2016eik}, \cite{Hamilton:2006az}, \cite{Kabat:2011rz}, \cite{Miyaji:2015fia}, \cite{Nakayama:2015mva} and \cite{Verlinde:2015qfa}.

\subsubsection*{Covariantization of MERA}
The choice to work on a constant time slice of the bulk space for computing explicit examples has been done mostly in order to obtain a simple comparison with MERA and entanglement density. However the geometric results are perfectly covariant, for any space-like slice $M_{sl}$, and we see no reason for not attaching to these the analogous physical meaning in terms of factorizability and entanglement. Hopefully this can be a guideline in fully understanding how to covariantize tensor networks, see for example \cite{Miyaji:2015fia} in this context.

\subsubsection*{Tripartite entanglement and beyond}
An obvious question is if it exists a tripartite entanglement version of our measure, or higher generalizations. A possible starting point would be to use relative entropy for multipartite states, $S(\rho_{A_1\cup A_2\cup A_3}\parallel \rho_{A_1}\otimes\rho_{A_2}\otimes\rho_{A_3})=S(A_1)+S(A_2)+S(A_3)-S(A_1\cup A_2\cup A_3)$, repeat the holographic computation in the appendix and use the result as an hint towards a possible geometric measure.

\subsubsection*{Higher dimensionality} 
The generalization to three and four boundary dimensions is an interesting but complicated geometrical problem, mostly because an higher dimensional surface has "more ways" to differ from an extremal one than a curve to differ from a geodesic. Analogously we would need a better understanding of these additional terms from the point of view of the CFT state. Nonetheless the main idea goes through, that an extremal surface is dual to a locally factorized state as the mutual information vanishes.

\subsubsection*{Relation with differential entropy}
An interesting geometric result from \cite{Balasubramanian:2013lsa} that received a quantum information theoretic interpretation in \cite{Czech:2014tva}, is that the length of a convex curve $\gamma$ can be expressed as the integral of the difference between the entanglement entropies of certain configurations of boundary intervals  $I(x)$, centred at $x$:
\be \label{dif}
\frac{\text{lenth}(\gamma)}{4G_N}=\int \Big(S(I(x))-S(I(x)\cap I(x-dx))\Big).
\ee
The shape of $\gamma$ in (\ref{dif}) determines the choice of sizes for $I(x)$ such that the corresponding Ryu-Takayanagi curves nicely cancel each other but for the infinitesimal part tangent to $\gamma$, that builds up the result.

Equation (\ref{dif}), whose right hand side received the name of differential entropy, has obvious similarities with (\ref{intlf}). Thus it is intriguing to imagine some connection between the two results: on one side a surface-state interpretation of (\ref{dif}) as some measure associated with the state $\ket{\psi}_{\gamma}$, on the other a boundary-CFT information theoretic understanding of (\ref{intlf}).

\subsubsection*{Computing Complexity}
As much as counting MERA disentanglers below a geodesic, bounded by an interval $\tilde{\gamma}_A$, should be interpreted as a measure of bipartite entanglement for $\rho_{\tilde{\gamma}_A}$, adding to the counting the number of coarse graining operations as well should produce a viable definition for complexity \footnote{I thank Giancarlo Camilo for proposing the idea.} for $\rho_{\tilde{\gamma}_A}$, with reference state the locally geodesic state $\rho_{\gamma_A}$. Analogously, by extending the region to include the full curve, we have a measure of complexity for $\ket{\psi}_{\gamma}$, with reference the boundary state with no real entanglement dual to a point, see \cite{Miyaji:2015yva} and \cite{Miyaji:2015fia}. As there is a coarse graining operation for each disentangler in the discrete MERA network and if we associate the same complexity weight for both operations, from the geometric point of view just by multiplying by two our result (\ref{diff2}) we would have a complexity measure. This idea resembles, and generalises to subsystems of a large class of CFT states, two other famous proposals for holographic complexity measures \cite{Brown:2015bva} and \cite{Stanford:2014jda}.

\subsubsection*{Quantum corrections}
To generalize our discussion to quantum bulk corrections is far from obvious but indeed a very interesting goal.
Quantum corrections to entanglement entropy have been discussed in various papers, see for example
\cite{Barrella:2013wja},  \cite{Engelhardt:2014gca}, \cite{Faulkner:2013yia}, \cite{Faulkner:2013ana}  and \cite{Prudenziati:2019pln}. 

The first thing that we notice is that the result of vanishing mutual information for a geodesic state goes through at least to order $O(1)$ in $G_N$. This because at this order quantum corrections to the entanglement entropy have been shown to assume the following structure, \cite{Faulkner:2013ana}:
\be 
S_{quantum}=S_{bulk}+\frac{\delta A}{4G_N}+\braket{\Delta S_{Wald-like}}+S_{counter},
\ee 
with $S_{bulk}$ referring to the bulk entanglement entropy with bulk entangling region given by the inside of the Ryu-Takayanagi curve and boundary entangling region, while the remaining terms are line integrals. Then, as the Ryu-Takayanagi curve coincides with the entangling region for a geodesic state, the latter terms cancel each other in computing the mutual information analogously to the classical  part, while $S_{bulk}$ trivially vanishes because its entangling region squeezes to zero. Because of this we still expect a geodesic state to be locally factorized in all the couples of subregions, either contiguous or not, at least up to order $O(G_N^0)$. What instead will not work is all the discussion on how measuring entanglement among any two contiguous regions implies $J(\gamma)$ to be a measure of entanglement along non-contiguous subregions as well. Because of this, disagreement with the entanglement density formalism is expected to show up when considering quantum bulk corrections. 

\section*{Acknowledgments}
I would like to thank Rafael Chavez for discussion, Tadashi Takayanagi for email correspondence and Aditya Mehra, Dmitry Melnikov, Filiberto Ares, Giancarlo Camilo, M\'at\'e Lencs\'es and Thiago Fleury for attending a seminar presenting a first draft of the paper and providing interesting feedback. This work has been done under financial support from the Brazilian ministries MCTI and MEC.

\appendix
\section{Mutual Information and geodesic curvature}
\begin{figure}
    \centering
    \includegraphics[width=0.7\textwidth]{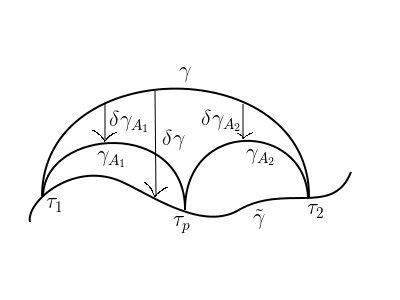}
    \caption{Holographic mutual information $I_{\tilde{\gamma}}(A_1:A_2)$.}
    \label{figuraMI}
\end{figure}

We consider the setup of figure \ref{figuraMI} 
to compute the mutual information $I_{\tilde{\gamma}}(A_1:A_2)$ for a state $\ket{\psi}_{\tilde{\gamma}}$ and two infinitesimal intervals $A_1,A_2$ with $\partial A_1=\{\tau_1,\tau_p\}$ and $\partial A_2=\{\tau_p,\tau_2\}$. The proper distance $d\tau^2=d\xi^a d\xi^b \delta_{ab}$ for a locally flat metric parametrizes both the curve $\tilde{\gamma}$ and the nearby Ryu-Takayanagi geodesics $\gamma_{A_1}, \gamma_{A_2}$ and $\gamma$ for the regions considered here. The definitions for the displacement vectors are:
\be
\gamma_{A_1}=\gamma+\delta \gamma_{A_1}  \;\;\;\; \gamma_{A_2}=\gamma+\delta \gamma_{A_2} \;\;\;\;\tilde{\gamma}=\gamma+\delta \gamma 
\ee
and their boundary conditions:
\be\label{bc}
\delta \gamma(\tau_1)=\delta \gamma(\tau_2)= \delta\gamma_{A_1}(\tau_1)=\delta\gamma_{A_2}(\tau_2)=0 \;\;\;\;\;\delta \gamma_{A_1}(\tau_p)=\delta \gamma_{A_2}(\tau_p)=\delta\gamma(\tau_p).         
\ee
The mutual information, from now on shortly indicated as $I$, is holographically computed as ($c=\hbar=1$):
\be
I=S_{\tilde{\gamma}}(A_1)+S_{\tilde{\gamma}}(A_2)-S_{\tilde{\gamma}}(A_1A_2)=\frac{1}{4G_N}\int_{\tau_1}^{\tau_p}\sqrt{\partial\gamma^{\mu}_{A_1}\partial\gamma^{\nu}_{A_1}g_{\mu\nu}}d\tau+
\ee
\[
+\frac{1}{4G_N}\int_{\tau_p}^{\tau_2}\sqrt{\partial\gamma^{\mu}_{A_2}\partial\gamma^{\nu}_{A_2}g_{\mu\nu}}d\tau- \frac{1}{4G_N}\int_{\tau_1}^{\tau_2}\sqrt{\partial\gamma^{\mu}\partial\gamma^{\nu}g_{\mu\nu}}d\tau
\]
and we denote $\partial=\frac{\partial}{\partial \tau}$. We first rescale $\tau$ so that $\partial\gamma^{\mu}\partial\gamma^{\nu}g_{\mu\nu}=v=1$, next we simplify the remaining two integrands expanding at first order in $\delta\gamma_{A_1}$ and $\delta\gamma_{A_2}$. This simplifies the mutual information to:
\be 
4G_NI=\int_{\tau_1}^{\tau_p}\partial\delta\gamma^{\mu}_{A_1}\partial\gamma^{\nu}g_{\mu\nu}d\tau
+\int_{\tau_p}^{\tau_2}\partial\delta\gamma^{\mu}_{A_2}\partial\gamma^{\nu}g_{\mu\nu}d\tau +O(\delta\gamma_{A_1,A_2}^2).
\ee 
We now perform integration by parts and use the boundary conditions (\ref{bc}) to obtain
\be 
4G_NI=-\int_{\tau_1}^{\tau_p}d\tau\left(\delta\gamma^{\mu}_{A_1}\partial^2\gamma^{\nu}g_{\mu\nu}+\delta\gamma^{\mu}_{A_1}\partial\gamma^{\nu}\partial g_{\mu\nu}\right)
-\int_{\tau_p}^{\tau_2}d\tau\left(\delta\gamma^{\mu}_{A_2}\partial^2\gamma^{\nu}g_{\mu\nu}+\delta\gamma^{\mu}_{A_2}\partial\gamma^{\nu}\partial g_{\mu\nu}\right).
\ee
Using the geodesic equation for $\gamma$
\[
\partial^2\gamma^{\mu}+\Gamma^{\mu}_{\rho\sigma}\partial\gamma^{\rho}\partial\gamma^{\sigma}=0,
\]
and the metricity for the metric $\bigtriangledown_{\tau}g_{\mu\nu}=0$ the expression further transforms to:
\be \label{int2}
4G_NI=-\int_{\tau_1}^{\tau_p}d\tau\delta\gamma^{\mu}_{A_1}\partial\gamma^{\rho}\partial\gamma^{\sigma}\Gamma^{\nu}_{\rho\mu}g_{\nu\sigma}-\int_{\tau_p}^{\tau_2}d\tau\delta\gamma^{\mu}_{A_2}\partial\gamma^{\rho}\partial\gamma^{\sigma}\Gamma^{\nu}_{\rho\mu}g_{\nu\sigma}.
\ee
Now we Taylor expand $\delta\gamma_{A_1}(\tau)$ to second order in $\tau-\tau_1$ 
\begin{align*} 
\delta\gamma_{A_1}(\tau)=\delta\gamma_{A_1}|_{\tau_1}+(\tau-\tau_1)\partial\delta\gamma_{A_1}|_{\tau_1}+\frac{(\tau-\tau_1)^2}{2}\partial^2\delta\gamma_{A_1}|_{\tau_1}+O((\tau-\tau_1)^3)
\end{align*}
and similarly $\delta\gamma_{A_2}(\tau)$ to second order in $\tau-\tau_2$. The constant term vanishes because of (\ref{bc}) while the second derivative can be transformed to a first derivative by application of the formula that constrains the infinitesimal displacement between two nearby geodesics, here $\gamma$ and either $\gamma_{A_1(A_2)}$,
\be\label{gd}
\partial^2\delta\gamma_{A_1(A_2)}^{\mu}=-\frac{\Gamma^{\mu}_{\nu\lambda}}{\partial\gamma^{\rho}}\delta\gamma_{A_1(A_2)}^{\rho}\partial\gamma^{\nu}\partial\gamma^{\lambda}
-2\Gamma^{\mu}_{\nu\lambda}\partial\gamma^{\nu}\partial\delta\gamma^{\lambda}_{A_1(A_2)},
\ee
and (\ref{bc}). Doing so we obtain
\be \label{int1}
\delta\gamma_{A_1}^{\mu}(\tau)=(\tau-\tau_1)\partial\delta\gamma_{A_1}^{\mu}|_{\tau_1}\left(\delta_{\rho}^{\mu}-(\tau-\tau_1)(\Gamma^{\mu}_{\nu\rho}\partial\gamma^{\nu})_{\tau_1}\right)
\ee
and analogously for $\delta\gamma_{A_2}$ around $\tau_2$. Once more we impose (\ref{bc}) 
\[
\delta\gamma_{A_1}^{\mu}(\tau_p)=(\tau_p-\tau_1)\partial\delta\gamma_{A_1}^{\mu}|_{\tau_1}\left(\delta_{\rho}^{\mu}-(\tau_p-\tau_1)(\Gamma^{\mu}_{\nu\rho}\partial\gamma^{\nu})_{\tau_1}\right) =\delta\gamma^{\mu}(\tau_p)
\]
with a similar equation for  $\delta\gamma_{A_2}$ replacing $\tau_1\leftrightarrow\tau_2$.
This permits us to solve for $\partial\delta\gamma_{A_1}|_{\tau_1}$ (resp. $\partial\delta\gamma_{A_2}|_{\tau_2}$) by inverting the matrix in between brackets above at first order in $(\tau_p-\tau_1)$ (resp. $(\tau_p-\tau_2)$):
\be 
\partial\delta\gamma_{A_1}^{\mu}|_{\tau_1}=\frac{\delta\gamma^{\rho}|_{\tau_p}}{(\tau_p-\tau_1)}\left(\delta_{\rho}^{\mu}+(\tau_p-\tau_1)(\Gamma^{\mu}_{\nu\rho}\partial\gamma^{\nu})_{\tau_1}\right)
\ee
and similarly for  $\partial\delta\gamma_{A_2}|_{\tau_2}$. Then we insert these results inside (\ref{int1}) (and its counterpart for $\delta\gamma_{A_1}^{\mu}(\tau)$) so that
\[
\delta\gamma_{A_1}^{\mu}(\tau)=\frac{(\tau-\tau_1)}{(\tau_p-\tau_1)}\left(\delta_{\lambda}^{\mu}-(\tau-\tau_p)(\Gamma^{\mu}_{\nu\lambda}\partial\gamma^{\nu})_{\tau_1}\right)\delta\gamma^{\lambda}|_{\tau_p}
\]
and again a similar story for $(A_1,\tau_1)\leftrightarrow (A_2,\tau_2)$.
Inserting back into the expression for the mutual information (\ref{int2}) we have
\be \label{int3}
4G_NI=-\int_{\tau_1}^{\tau_p}d\tau\frac{(\tau-\tau_1)}{(\tau_p-\tau_1)}\left(\delta_{\lambda}^{\mu}-(\tau-\tau_p)(\Gamma^{\mu}_{\nu\lambda}\partial\gamma^{\nu})_{\tau_1}\right)\delta\gamma^{\lambda}|_{\tau_p}\partial\gamma^{\rho}\partial\gamma^{\sigma}\Gamma^{\nu}_{\rho\mu}g_{\nu\sigma}-
\ee
\[
-\int_{\tau_p}^{\tau_2}d\tau\frac{(\tau-\tau_2)}{(\tau_p-\tau_2)}\left(\delta_{\lambda}^{\mu}-(\tau-\tau_p)(\Gamma^{\mu}_{\nu\lambda}\partial\gamma^{\nu})_{\tau_2}\right)\delta\gamma^{\lambda}|_{\tau_p}\partial\gamma^{\rho}\partial\gamma^{\sigma}\Gamma^{\nu}_{\rho\mu}g_{\nu\sigma}.
\]
The next step is to expand $\delta\gamma^{\lambda}|_{\tau_p}$ around $\delta\gamma^{\lambda}|_{\tau_1}$ and $\delta\gamma^{\lambda}|_{\tau_2}$ in the two integrands, again at second order in the $\tau$ variation. The key point here is that the displacement $\delta\gamma$ is not between two geodesics but between a geodesic $\gamma$ and a nearby generic convex curve $\tilde{\gamma}$, so equation (\ref{gd}) acquires a new term which is the vector $k_c^{\mu}$ defined as $k_c^{\mu}(\tilde{\gamma})=\bigtriangledown_{\tau}\partial_{\tau}\tilde{\gamma}^{\mu}$:
\be\label{gc}
\partial^2\delta\gamma^{\mu}=-\partial_{\rho}\Gamma^{\mu}_{\nu\lambda}\delta\gamma^{\rho}\partial\gamma^{\nu}
\partial\gamma^{\lambda}-2\Gamma^{\mu}_{\nu\lambda}\partial\gamma^{\nu}\partial\delta\gamma^{\lambda}+k_c^{\mu}(\tilde{\gamma}).
\ee
A few remarks: first everything in the above equation is along $\gamma$ but $k_c^{\mu}$ that is evaluated on $\tilde{\gamma}$. Second there should be a normalization $\parallel\partial\tilde{\gamma}\parallel^2$ multiplying $k_c^{\mu}(\tilde{\gamma})$, as the definition assumes a unit tangent vector; nonetheless the normalization is $\parallel\partial\tilde{\gamma}\parallel=1+O(\delta\gamma)$, so being $k_c$ already of order $O(\delta\gamma)$ we can discard this correction. 

We then essentially repeat the previous steps for the evaluation of the second order expansion of  $\delta\gamma^{\lambda}(\tau_p)$; first we expand it at second order around $\tau_1$ (resp. $\tau_2$) and use the boundary conditions $\delta\gamma|_{\tau_2}=0$ (resp. $\delta\gamma|_{\tau_1}=0$) from (\ref{bc}) to fix the constant term and the derivatives  obtaining
\be \label{int4}
\delta\gamma|_{\tau_1}=0 \;\;\;\;\;\partial\delta\gamma|_{\tau_1}= -\frac{1}{2}(\tau_2-\tau_1)\partial^2\delta\gamma|^{TM_{sl}}_{\tau_1} \;\;\;\;\;\partial^2\delta\gamma|^{NM_{sl}}_{\tau_1}=0
\ee
(and similarly in $\tau_2$), where we have indicated as $TM_{sl}$ and $NM_{sl}$ the projection along the tangent and the normal to the space-like slice $M_{sl}$ to which $\tilde{\gamma}$ belongs. After simple algebra it produces:
\be \label{int6}
\delta\gamma(\tau_p)= \frac{1}{2}(\tau_p-\tau_1)(\tau_p-\tau_2)\partial^2\delta\gamma|_{\tau_1}^{TM_{sl}}
\ee
together with the equivalent equation from $\partial^2\delta\gamma|_{\tau_1}\leftrightarrow \partial^2\delta\gamma|_{\tau_2}$ to be used in the other integrand \footnote{Incidentally $\partial^2\delta\gamma|_{\tau_1}=\partial^2\delta\gamma|_{\tau_2}$ because of the form of (\ref{int6}) and (\ref{bc}), but we will still keep the two indices for clarity.}. With this expressions at hands we can use (\ref{gc}) in $\tau_1$ (resp. $\tau_2$) plus (\ref{int4}) to obtain an equation for the second derivative \footnote{See footnote \ref{ftn} for reference on the definition of the various $k$s.}:
\be 
\partial^2\delta\gamma^{\mu}|_{\tau_1}^{TM_{sl}}\left(\delta_{\mu}^{\nu}-\left(\Gamma^{\mu}_{\nu\lambda}\partial\gamma^{\nu}\right)_{\tau_1}(\tau_2-\tau_1)\right)=k_c^{\mu}(\tilde{\gamma})|_{\tau_1}^{TM_{sl}}=k^{\mu}(\tilde{\gamma})|_{\tau_1}
\ee
that inverted at first order and inserted into (\ref{int6}) produces:
\be  \label{int5}
\delta\gamma^{\mu}(\tau_p)=\frac{1}{2}(\tau_p-\tau_1)(\tau_p-\tau_2)k^{\nu}(\tilde{\gamma})|_{\tau_1}\left(\delta_{\mu}^{\nu}+\left(\Gamma^{\mu}_{\nu\lambda}\partial\gamma^{\nu}\right)_{\tau_1}(\tau_2-\tau_1)\right)
\ee
and again we have a similar equation by exchanging $\tau_1\leftrightarrow \tau_2$. Plugging (\ref{int5}) inside (\ref{int3}) gives, at second order in the $\tau$ variation:
\be \label{mi2}
4G_NI=\int_{\tau_1}^{\tau_p}d\tau\frac{(\tau-\tau_1)(\tau_2-\tau_p)}{2}\partial\gamma^{\theta}\partial\gamma^{\sigma}
\Gamma^{\xi}_{\theta\mu}g_{\xi\sigma}k^{\nu}(\tilde{\gamma})|_{\tau_1}\left(\delta_{\nu}^{\mu}+[(\tau_p-\tau)+(\tau_2-\tau_1)](\Gamma^{\mu}_{\rho\nu}\partial\gamma^{\rho})_{\tau_1}\right)
\ee
\[
+\int_{\tau_p}^{\tau_2}d\tau\frac{(\tau_2-\tau)(\tau_p-\tau_1)}{2}\partial\gamma^{\theta}\partial\gamma^{\sigma}\Gamma^{\xi}_{\theta\mu}g_{\xi\sigma}k^{\nu}(\tilde{\gamma})|_{\tau_2}\left(\delta_{\nu}^{\mu}+[(\tau_p-\tau)+(\tau_1-\tau_2)](\Gamma^{\mu}_{\rho\nu}\partial\gamma^{\rho})_{\tau_2}\right).
\]
Finally we fix $\Delta\tau=(\tau_p-\tau_1)=(\tau_2-\tau_p)$, compute the double derivative $\partial^2_{\Delta\tau}I$ and expand at first order in $\Delta\tau$. Rewriting
\be\label{te}
\partial\gamma^{\theta}\partial\gamma^{\sigma}\Gamma^{\xi}_{\theta\mu}g_{\xi\sigma}k^{\mu}(\tilde{\gamma})=\partial\gamma^{\theta}\partial\gamma^{\sigma}\partial_{\nu} g_{\theta\sigma}k^{\nu}(\tilde{\gamma})
\ee
we finally obtain
\be \label{hj}
J_S(\tilde{\gamma})|_{\tau_p}=\partial_{\Delta\tau}^2I|_{\tau_p}=\frac{3\Delta\tau}{4G_N}\left(\partial\tilde{\gamma}^{\theta}\partial\tilde{\gamma}^{\sigma}\partial_{\nu} g_{\theta\sigma}k^{\nu}\right)|_{\tau_p}+O(\Delta\tau^3).
\ee
Note that formally the the term $\partial\gamma^{\theta}\partial\gamma^{\sigma}\partial_{\nu} g_{\theta\sigma}$ from (\ref{te}) should be computed along the geodesic $\gamma$, but at leading order in $\Delta\tau$ this just coincides with the same quantity along the curve $\tilde{\gamma}$, so this justifies equation (\ref{hj}) where everything is along $\tilde{\gamma}$.

\end{document}